\documentclass[conference]{IEEEtran}
\IEEEoverridecommandlockouts
\usepackage{cite}
\usepackage{overpic}
\usepackage{amsmath,amssymb,amsfonts}
\usepackage{graphicx}
\usepackage{textcomp}
\usepackage{xcolor}
\usepackage{float}
\usepackage{amsthm}
\usepackage{graphicx}
\usepackage{epstopdf}
\usepackage{amsmath,bm}
\usepackage{amsfonts}
\usepackage{amssymb}
\usepackage{color}
\usepackage{multirow}
\usepackage{multicol}
\usepackage{soul,xcolor}
\usepackage{algorithm}% http://ctan.org/pkg/algorithms
\usepackage{algpseudocode}%
\usepackage{float}
\usepackage{graphicx}    % for including graphics
\usepackage{subcaption}  % for subfigures and subcaptions

\theoremstyle{plain}

\newtheorem{remark}{Remark}

\newcommand{\vect}[1]{\mathbf{#1}}

\def\diag{\mathrm{diag}}

\def\Htran{\mbox{\tiny $\mathrm{H}$}}
\def\Ttran{\mbox{\tiny $\mathrm{T}$}}
\def\imagunit{\mathsf{j}} % Imaginary number

\newcommand{\argmax}[1]{{\underset{{#1}}{\mathrm{arg\,max}}}}
\newcommand{\argmin}[1]{{\underset{{#1}}{\mathrm{arg\,min}}}}

\begin{document}

\title{Parametric Channel Estimation for RIS-Assisted Wideband Systems   \vspace{-0.2cm}
%\thanks{This work was supported by the  FFL18-0277 grant from the Swedish Foundation for Strategic Research.}
}

\author{\IEEEauthorblockN{Alva Kosasih$^*$, \"Ozlem Tu\u{g}fe Demir$^{\dagger}$, and Emil Bj{\"o}rnson$^*$ \thanks{This work was supported by the FFL18-0277 grant from the Swedish Foundation for Strategic Research.} }\\
\IEEEauthorblockA{{$^*$Department of Computer Science, KTH Royal Institute of Technology, Kista, Sweden
		}\\{$^\dagger$Department of Electrical-Electronics Engineering, TOBB University of Economics and Technology, Ankara, Turkey
		} \\ 
		{Email: kosasih@kth.se, ozlemtugfedemir@etu.edu.tr, emilbjo@kth.se\vspace{-0.4cm}}
}
% make the title area
}

\maketitle

\begin{abstract}
A reconfigurable intelligent surface (RIS) alters the reflection of incoming signals based on the phase-shift configuration assigned to its elements. This feature can be used to improve the signal strength for user equipments (UEs), expand coverage, and enhance spectral efficiency in wideband communication systems. 
Having accurate channel state information (CSI) is indispensable to realize the full potential of RIS-aided wideband systems. Unfortunately, CSI is challenging to acquire due to the passive nature of the RIS elements, which cannot perform transmit/receive signal processing.
Recently, a parametric maximum likelihood (ML) channel estimator has been developed and demonstrated excellent estimation accuracy. However, this estimator is designed for narrowband systems with no non-line-of-sight (NLOS) paths. In this paper, we develop a novel parametric ML channel estimator for RIS-assisted wideband systems, which can handle line-of-sight (LOS) paths in the base station (BS)-RIS and RIS-UE links as well as NLOS paths between the UE, BS, and RIS. We leverage the reduced subspace representation induced by the array geometry to suppress noise in unused dimensions, enabling accurate estimation of the NLOS paths. Our proposed algorithm demonstrates superior estimation performance for the BS-UE and RIS-UE channels, outperforming the existing ML channel estimator. 
\end{abstract}


\section{Introduction}
\label{S_Intro}

Reconfigurable intelligent surface (RIS)-aided communication stands out as a key new ingredient of next-generation wireless systems, offering a way to enhance propagation conditions between a base station (BS) and user equipment (UE) \cite{2019_Huang_TWC,RIS_Renzo_JSAC}. 
An RIS, consisting of low-cost passive elements, reflects impinging signals with a reflection pattern controlled through adjustable phase-shift induction in each RIS element. This enables shaping the reflected wavefront, such as steering it as a beam toward the intended UE. The primary challenge in such a scenario is the channel estimation from/to the RIS elements. Since the RIS reflects signal instead of receiving, channel estimation relies on transmitting  pilot signals and modifying the phase-shift configuration repeatedly to enable the receiver to estimate the cascaded channel.

The previous methods proposed for RIS channel estimations can be broadly classified into non-parametric and parametric approaches.  In non-parametric approaches, least squares (LS) \cite{2020_Jensen_ICASSP} and linear minimum mean square error (LMMSE) estimators \cite{2021_Kundu_IEEE} are commonly used, with their primary difference lying in the utilization of deterministic or stochastic models, respectively. The use of a stochastic model in LMMSE estimation provides additional gains compared to LS in low SNR scenarios, albeit at the cost of requiring statistical knowledge in the form of the spatial correlation matrix, which might be hard to acquire in practice. Nevertheless, both LS and LMMSE are considered easy to implement and provide decent estimation performance. Additionally, compressive sensing approaches have been proposed to solve the RIS channel estimation problem. First, a sparse transformation of the channel is needed. Then, methods such as iterative hard thresholding (IHT) \cite{2013_Carrillo_TSP}, orthogonal matching pursuit (OMP) \cite{2007_Tropp_TIT}, alternating direction method of multipliers (ADMM) \cite{2020_Yu_Access}, and  approximate message passing (AMP) \cite{2014_Parker_TSP} can be used to estimate the transformed channel matrix. However, the sparse transformation involves the utilization of a discrete dictionary that results in huge complexity when a high resolution is needed \cite{2022_Swindlehurst_IEEE}. 

The main reason for spatial sparsity is the existence of a dominant line-of-sight (LOS) path. Instead of discovering it by compressive sensing, a parametric estimator can be designed to estimate it directly. A parametric maximum likelihood  estimator (MLE) was introduced in \cite{2021_Wang_TWC,2022_Björnson_Asilomar} for scenarios with unknown free-space far-field LOS channels to and from the RIS. A similar parametric beam training procedure was described in \cite{2021_Ning_TVT}. The case with near-field LOS channels was recently considered in \cite{Haghshenas2024a}, where the main focus was to obtain accurate MLE with only a few pilots. This work confirmed the superiority of the parametric MLE compared to the non-parametric LS estimator. However, the above works are limited to narrowband free-space LOS channels. 
While it can be argued that the narrowband assumption is sufficient for RIS phase-shift design when there are only LOS paths to and from the RIS, practical terrestrial systems will contain non-negligible non-LOS (NLOS) paths that result in wideband channel conditions.
In the presence of a significant NLOS component, the existing MLEs are mismatched and cannot give any performance guarantees. Hence, there is a research gap in developing a parametric MLE for RIS-assisted wideband systems.

In this paper, we aim to develop the first parametric MLE for RIS-assisted wideband systems with both LOS and NLOS paths. When estimating the NLOS paths, we propose to exploit the reduced subspace representation induced by the array geometry \cite{demir2022channel} to efficiently suppress noise in unexisting channel dimensions. This approach aims to outperform the LS estimator of the NLOS components without relying on spatial correlation knowledge. Our proposed estimator can be viewed as a generalization of the parametric MLE in \cite{2022_Björnson_Asilomar,Haghshenas2024a} for RIS wideband systems.  
We will demonstrate the estimation performance benefits of the proposed estimator, compared to prior work, through numerical simulations.

\section{System and Channel Modeling}
\label{S_SysMod}

\begin{figure}
    \centering
    \begin{overpic}[scale=0.6]{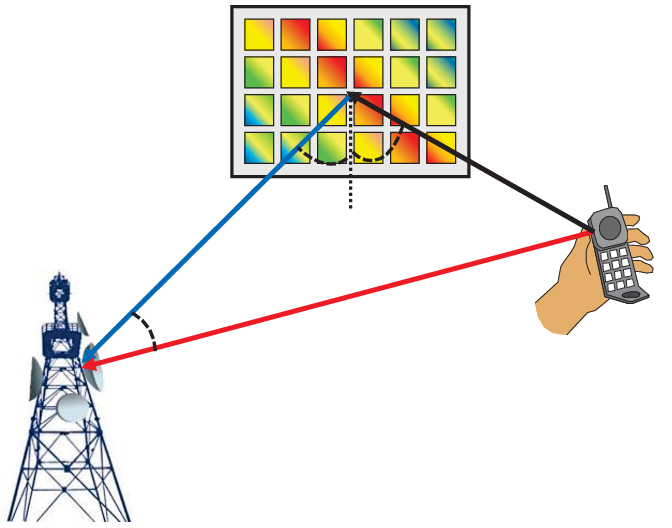}
    \put(22,45){$\vect{H}[s]$}
    \put(76,53){$\vect{g}[s]$}
    \put(50,30){$\vect{d}[s]$}
    \put(41,49){$\bar{\varphi}_{\rm RIS}$}
    \put(58,49){$\varphi$}
    \put(23,31){$\varphi_{\rm UE}$}
    \end{overpic}
    \caption{System model with $M$-antenna BS, $N$-element RIS, and a single-antenna UE.}
    \label{F_illus}
\end{figure}
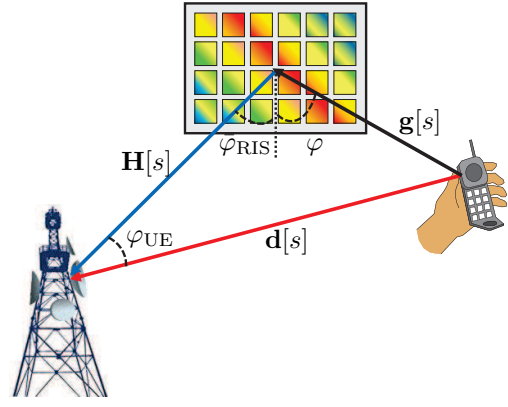

We consider a RIS-assisted communication between a BS and UE as illustrated in Fig.~\ref{F_illus}.
The UE is equipped with a single antenna, while the BS is equipped with $M$ antennas, geometrically shaped in a uniform planar array (UPA) with $M_{\rm H}$ and $M_{\rm V}$ elements per row and column, respectively. The RIS has $N$ elements, geometrically shaped in a UPA with $N_{\rm H}$  and $N_{\rm V}$ elements per row and column, respectively. We consider a wideband channel with $\bar{S}$ subcarriers, whereof $S$ subcarriers are used for pilot transmissions. 
The pilots are distributed evenly across subcarriers, with sufficient spacing so that the correlation between any two considered subcarriers is negligible.\footnote{This holds if the $S$ is equal to the number of taps in the underlying time-domain channel.}
To this end, we focus solely on the subcarriers designated for pilot transmissions, without loss of generality.

The channel between the BS and RIS is known at the BS. This is motivated by the permanent locations of the BS and RIS and intermittent training.  The channel between BS and RIS  is denoted by $\vect{H}[s]\in \mathbb{C}^{M \times N}$ at the subcarrier $s \in \{1,\dots, S\}$ and can be expressed as
\begin{align}
 \vect{H}[s] &= \sqrt{\bar{\beta}}e^{\imagunit \bar{\phi}}\vect{\bar{a}}\left(\bar{\varphi}_{\rm BS},\bar{\theta}_{\rm BS}\right)\vect{a}^{\Ttran}\left(\bar{\varphi}_{\rm RIS},\bar{\theta}_{\rm RIS}\right) + \breve{\vect{H}}[s],
\end{align}
where $\bar{\beta}$ is the LOS channel gain and $\bar{\phi}$ is the phase-shift of the LOS part. The $\vect{\bar{a}}(\bar{\varphi}_{\rm BS},\bar{\theta}_{\rm BS})$ and $\vect{a}(\bar{\varphi}_{\rm RIS},\bar{\theta}_{\rm RIS})$ are the UPA response vectors at the pair of azimuth and elevation angles seen from the BS and the RIS, respectively.  The last term $\breve{\vect{H}}[s]$ is the NLOS part, which can be expressed as
\begin{align}\label{eq:NLOS_ch_model}
\breve{\vect{H}}[s] = \sum_{l=1}^L \alpha_{s,l}\vect{\bar{a}}\left(\bar{\varphi}_{{\rm BS}}^{(l)}, \bar{\theta}_{{\rm BS}}^{(l)}\right)\vect{a}^{\Ttran}\left(\bar{\varphi}_{{\rm RIS}}^{(l)},\bar{\theta}_{{\rm RIS}}^{(l)}\right)
\end{align}
where $L$ is the number of NLOS propagation clusters and $\alpha_{s,l}$ is the relative complex attenuation factor of the $l$-th cluster at subcarrier $s$, modelled by a complex independent and identically distributed (i.i.d.) Gaussian distribution.
The pairs of azimuth and elevation angles, $(\bar{\varphi}_{{\rm BS}}^{(l)}, \bar{\theta}_{{\rm BS}}^{(l)})$ and $(\bar{\varphi}_{{\rm RIS}}^{(l)}, \bar{\theta}_{{\rm RIS}}^{(l)})$, are the corresponding angles seen by the BS and RIS at the $l$-th cluster, respectively.

The channel between the RIS and UE  at subcarrier $s$ is denoted by $\vect{g}[s]\in \mathbb{C}^N$ and it is unknown. This channel is also modeled similarly, i.e., 
it consists of  LOS and NLOS parts, expressed as follows:
\begin{align}\label{eq_user_channel}
    \vect{g}[s] = \sqrt{\beta} e^{\imagunit \phi}\vect{a}(\varphi)+\breve{\vect{g}}[s],
\end{align}
where the LOS channel gain $\beta$, phase-shift  $\phi$, and  the array response vector $\vect{a}(\varphi)$ from the UE to the RIS with an angle-of-arrival (AoA) $\varphi$ are unknown. 
To simplify the analysis, we assume the UE is positioned at the same elevation as the RIS, thus, $\theta=0$.
The first term in \eqref{eq_user_channel} belongs to the LOS component while the second term belongs to the NLOS component. 
The NLOS paths are spatially correlated, but the corresponding spatial statistics or correlation matrix is unknown. 
 As illustrated in Fig.~\ref{F_illus}, there is also an NLOS channel between the BS and UE, denoted by $\vect{d}[s]\in \mathbb{C}^M$ at subcarrier $s$. We assume the LOS path is blocked and the spatial statistics is unknown. 
 
To estimate channels, the UE sends pilots across $S$ subcarriers. Since we do not know the distributions of the channels $\breve{\vect{g}}[s]$ and $\vect{d}[s]$, we treat them as unknown but deterministic vectors. As shown in \cite{demir2022channel}, there is a reduced subspace in which such a channel can reside induced by the array geometry and element spacing in a planar array, and this effect is more pronounced when the element spacing decreases. Hence, we know that $\breve{\vect{g}}[s]$ and $\vect{d}[s]$ exist in a subspace spanned by the columns of semi-unitary matrices $\vect{U}_{g}\in \mathbb{C}^{N\times r_g}$ and $\vect{U}_d\in \mathbb{C}^{M \times r_d}$, respectively.  The ranks of these matrices are $r_g$ and $r_d$, respectively. These reduced-subspace matrices solely depend on array geometry, cover all possible channel dimensions (for any possible spatial correlation matrix), and can be computed based on \cite[Eq.~(8)]{demir2022channel}. Accordingly, we can write the channel components as
\begin{align}
    &\breve{\vect{g}}[s] = \vect{U}_{g} \vect{x}_g[s], \quad \vect{d}[s] =  \vect{U}_d \vect{x}_d[s],
\end{align}
where  $\vect{x}_g[s]$ and $\vect{x}_d[s]$  are the reduced-subspace components of the channels, which are unknown and treated as deterministic.  If the UE sends the pilot symbol $\sqrt{P}$ at each pilot subcarrier, the received signal on subcarrier $s$ is
\begin{align}\label{eq_rec_signal}
    \vect{y}[s] &= \vect{d}[s]\sqrt{P}+\vect{H}[s]\vect{\Phi}\vect{g}[s]\sqrt{P} + \vect{n}[s] \nonumber \\
    & = \vect{U}_d\vect{x}_d[s]\sqrt{P}+ \sqrt{\beta}e^{\imagunit\phi } \vect{H}[s]\vect{\Phi}\vect{a}(\varphi)\sqrt{P} \nonumber \\  
    &\hspace{1em} + \vect{H}[s]\vect{\Phi}\vect{U}_g\vect{x}_g[s]\sqrt{P}+\vect{n}[s], \quad s=1,\ldots,S,
\end{align}
where $\vect{\Phi} = \diag(e^{\imagunit\theta_{1}},\ldots, e^{\imagunit\theta_{N}})$ is the configuration of the RIS during pilot transmission, $\vect{n}[s] \sim \mathcal{N}(\vect{0},\sigma^2 \vect{I}_M)$ is the additive white Gaussian noise (AWGN) at subcarrier $s$, and $\sigma^2$ is the noise variance. In contrast to previous works \cite{2022_Björnson_Asilomar,Haghshenas2024a}, which require multiple pilots to be transmitted using different RIS configurations, we consider only a single RIS configuration. The reason that we only need a single RIS configuration is that we exploit multiple BS antennas and subcarrier diversity, which will be further explained in the following section.

\section{Parametric Maximum Likelihood Estimation}
\label{S_MLE}

We can write the probability density function  (PDF) of the received pilot signals in \eqref{eq_rec_signal} given the UE's channel parameters as
 \begin{align}
     &f\left(\left\{\vect{y}[s]\right\}_{s=1}^S\big|\varphi,\beta ,\phi ,\left\{\vect{x}_g[s],\vect{x}_d[s]\right\}_{s=1}^S\right) \nonumber\\
     &= \prod_{s=1}^S\frac{1}{\pi \sigma^2} e^{- \frac{ \left\| \vect{y}[s] - \vect{U}_d  \vect{x}_d[s] \sqrt{P} - {\vect{c}_g[s]}   \right\|^2}{\sigma^2}},
 \end{align}
 where ${\vect{c}_g[s]} = \sqrt{\beta } e^{\imagunit \phi  } \vect{H}[s]  \vect{\Phi} \vect{a}(\varphi) \sqrt{P} + \vect{H}[s] \vect{\Phi}  \vect{U}_g \vect{x}_g[s]\sqrt{P}  $ is a function of the array response vector to the UE $\vect{a}(\varphi)$.
 Considering all subcarrier indexes $1,\dots,S$, we can  write the maximum likelihood estimation (MLE) problem as in \eqref{eq_MLE} at the top of the next page.
\begin{figure*}
\begin{align}\label{eq_MLE}
\nonumber
& \left \{\hat{\beta} ,\hat{\phi} ,\hat{\varphi},\hat{\vect{x}}_g[s], \hat{\vect{x}}_d[s]\right\} =  \argmin{\substack{\beta ,\phi , \varphi,\\ \vect{x}_g[s], \vect{x}_d[s]}} \sum_{s=1}^S \left\| \vect{y}[s] - \vect{U}_d  \vect{x}_d[s] \sqrt{P} -\left(  \underbrace{\sqrt{\beta } e^{\imagunit \phi  } \vect{H}[s]  \vect{\Phi} \vect{a}(\varphi) \sqrt{P} + \vect{H}[s] \vect{\Phi}  \vect{U}_g \vect{x}_g[s]\sqrt{P}  }_{\vect{c}_g[s]} \right)   \right\|^2, 
\nonumber \\ 
 &=\argmin{\substack{\beta ,\phi , \varphi,\\ \vect{x}_g[s], \vect{x}_d[s]}} \sum_{s=1}^S \big(\vect{y}^{\Htran}[s] \vect{y}[s] + P \vect{x}_d^{\Htran}[s] \vect{x}_d[s] + \vect{c}_g^{\Htran}[s]\vect{c}_g[s]
 -2 \sqrt{P}\Re\left(  \vect{x}_d^{\Htran}[s]\vect{U}_d^{\Htran}\vect{y}[s] \right) + 2 \sqrt{P} \Re\left( \vect{x}_d^{\Htran}[s] \vect{U}_d^{\Htran} \vect{c}_g[s]\right)   -2 \Re\left( \vect{y}^{\Htran}[s]  \vect{c}_g[s]\right)\big).  
 \end{align}
\hrulefill
\end{figure*}

We begin by estimating  ${\vect{x}}_d[s]$ from the MLE problem in \eqref{eq_MLE}. More specifically, we first simplify the problem by ignoring all terms that are independent of ${\vect{x}}_d[s]$, since they will not change the result of the minimization. We then obtain
\begin{multline}
     \argmin{{\vect{x}}_d[s]}    \sum_{s=1}^S \Big( P \vect{x}_d^{\Htran}[s] \vect{x}_d[s] \\ 
     -2 \sqrt{P}\Re\Big(  \vect{x}_d^{\Htran}[s] \left(  \vect{U}_d^{\Htran}\vect{y}[s]  - \vect{U}_d^{\Htran} \vect{c}_g[s] \right)\Big)\Big).
\end{multline}
This is a quadratic function minimization problem and it is straightforward to show that it is minimized by
\begin{equation}\label{eq_x_d}
    \hat{\vect{x}}_d[s] = \frac{1}{\sqrt{P}} \left( \vect{U}_d ^{\Htran}\vect{y}[s] -\vect{U}_d^{\Htran}\vect{c}_g[s]  \right).
\end{equation}
Substituting \eqref{eq_x_d} into \eqref{eq_MLE} gives us the simplified MLE problem stated in \eqref{eq_MLE_4vars} at the top of the next page.
\begin{figure*}
\begin{align}\label{eq_MLE_4vars}
& \left \{\hat{\beta} ,\hat{\phi} ,\hat{\varphi},\hat{\vect{x}}_g[s]\right\} =
 \argmin{\beta ,\phi ,\varphi,\vect{{x}}_g[s]} \quad \sum_{s=1}^S\Bigg(\vect{y}^{\Htran}[s] \vect{y}[s] +    \left( \vect{U}_d ^{\Htran}\vect{y}[s]  -  \vect{U}_d^{\Htran}\vect{c}_g[s]  \right) ^{\Htran}     \left( \vect{U}_d ^{\Htran}\vect{y}[s]  - \vect{U}_d^{\Htran}\vect{c}_g[s]  \right)  + \vect{c}_g^{\Htran}[s]\vect{c}_g[s] \nonumber \\
&\quad-2  \Re\left(\vect{y}^{\Htran}[s] \vect{U}_d   \left( \vect{U}_d ^{\Htran}\vect{y}[s] 
 - \vect{U}_d^{\Htran}\vect{c}_g[s]  \right)  \right) + 2  \Re\left( \vect{c}_g^{\Htran}[s]\vect{U}_d 
 \left( \vect{U}_d ^{\Htran}\vect{y}[s]  - \vect{U}_d^{\Htran}\vect{c}_g[s] \right)  \right)  
  -2\Re\left( \vect{y}^{\Htran}[s]  \vect{c}_g[s]\right) \Bigg),
  \notag \\
  &= \argmin{\beta ,\phi ,\varphi,\vect{{x}}_g[s]} \quad \sum_{s=1}^S\Bigg( \vect{y}^{\Htran}[s] \left(\vect{I}_M - \vect{U}_d \vect{U}_d^{\Htran} \right ) \vect{y}[s]  + \vect{c}_g^{\Htran}[s]  \left (\vect{I}_M - \vect{U}_d \vect{U}_d^{\Htran} \right ) \vect{c}_g[s]  - 2 \Re\left( \vect{y}^{\Htran}[s] \left (\vect{I}_M-\vect{U}_d \vect{U}_d^{\Htran}\right )  \vect{c}_g[s]\right )\Bigg),   \nonumber\\
  & = \argmin{\substack{\beta ,\phi , \\ \varphi,\vect{{x}}_g[s]}} \quad \sum_{s=1}^S\Bigg( P {\beta}  \vect{a}^{\Htran}(\varphi) \overline{\vect{H}}^{\Htran}[s] (\vect{I}_M - \vect{U}_d\vect{U}_d^{\Htran}) \overline{\vect{H}}[s] \vect{a}(\varphi) + P \vect{x}_g^{\Htran}[s]  \vect{U}_g^{\Htran}  \overline{\vect{H}}^{\Htran}[s] (\vect{I}_M - \vect{U}_d \vect{U}_d^{\Htran}) \overline{\vect{H}}[s] \vect{U}_g \vect{x}_g[s] \nonumber\\
  & \hspace{6em} + 2P \sqrt{\beta } \Re\left( e^{-\imagunit \phi } \vect{a}^{\Htran}(\varphi) \overline{\vect{H}}^{\Htran}[s]  (\vect{I}_M - \vect{U}_d \vect{U}_d^{\Htran}) \overline{\vect{H}}[s]  \vect{U}_g \vect{x}_g[s]\right)  \nonumber\\ 
  & \hspace{6em}  - 2 \sqrt{P}\sqrt{\beta }\Re \left( \vect{y}^{\Htran}[s]  (\vect{I}_M-\vect{U}_d \vect{U}_d^{\Htran} )  e^{\imagunit \phi  } \overline{\vect{H}} [s]\vect{a}(\varphi) \right) - 2\sqrt{P}\Re \left(\vect{y}^{\Htran}[s](\vect{I}_M-\vect{U}_d\vect{U}_d^{\Htran})\overline{\vect{H}}[s]  \vect{U}_g \vect{x}_g[s] \right)\Bigg). 
\end{align}
\hrulefill
\end{figure*}
We define $\overline{\vect{H}}[s] \triangleq \vect{H}[s]\vect{\Phi}$. Minimizing the MLE objective with respect to $\vect{x}_g[s]$ is equivalent to the following problem:
\begin{align}\label{eq:min_xg_init}
   & \argmin{\vect{x}_g[s]} \sum_{s=1}^S \Big(  P \vect{x}_g^{\Htran}[s]  \vect{U}_g^{\Htran}   \overline{\vect{H}}^{\Htran}[s]  (\vect{I}_M - \vect{U}_d \vect{U}_d^{\Htran}) \overline{\vect{H}}[s] \vect{U}_g \vect{x}_g[s] \nonumber \\ 
  &  + 2  P \sqrt{\beta } \Re \left(e^{-\imagunit \phi } \vect{a}^{\Htran}(\varphi) \overline{\vect{H}}^{\Htran}[s] (\vect{I}_M - \vect{U}_d\vect{U}_d^{\Htran}) \overline{\vect{H}}[s] \vect{U}_g \vect{x}_g[s]   \right) \nonumber \\
  &- 2 \sqrt{P} \Re \left(  \vect{y}^{\Htran}[s]  (\vect{I}_M-\vect{U}_d \vect{U}_d^{\Htran}) \overline{\vect{H}}[s] \vect{U}_g \vect{x}_g[s] \right) \Big).
\end{align}
By defining the notation
\begin{align}\label{eq_Ag}
  &  \vect{A}_g[s] \triangleq  (\vect{I}_M-\vect{U}_d\vect{U}_d^{\Htran})\overline{\vect{H}}[s]\vect{U}_g, 
\end{align}
we can rewrite the minimization problem  with respect to $\vect{x}_g[s]$ as
\begin{multline}\label{eq_MLE_xg}
\argmin{\vect{{x}}_g[s] } \sum_{s=1}^S \Bigg( \vect{x}_g^{\Htran}[s]\vect{A}_g^{\Htran}[s]\vect{A}_g[s] \vect{x}_g[s]  - 2 \Re \bigg( \frac{1}{\sqrt{P}}  \vect{y}^{\Htran}[s]  \\ \vect{A}_g[s] \vect{x}_g[s]  - \sqrt{\beta }  e^{-\imagunit \phi } \vect{a}^{\Htran}(\varphi) \overline{\vect{H}}^{\Htran}[s] \vect{A}_g[s] \vect{x}_g[s] \bigg) \Bigg),
\end{multline}
where we utilized the fact that $(\vect{I}_M-\vect{U}_d\vect{U}_d^{\Htran})(\vect{I}_M-\vect{U}_d\vect{U}_d^{\Htran})= \vect{I}_M-\vect{U}_d\vect{U}_d^{\Htran}$ because $\vect{U}_d$ is a semi-unitary matrix.
Since we have quadratic minimization problem with respect to \eqref{eq_MLE_xg}, it is now straightforward to obtain the estimate of $\vect{x}_g[s]$ as
\begin{multline}\label{eq_pinv}
  \hat{\vect{x}}_g[s] = \left(\vect{A}_g^{\Htran}[s]\vect{A}_g[s]\right)^{\dagger} \vect{A}_g^{\Htran}[s] \\
  \times\Bigg(\frac{1}{\sqrt{P}}\vect{y}[s]  -\sqrt{\beta }e^{\imagunit\phi }\overline{\vect{H}}[s]\vect{a}(\varphi)  \Bigg).
\end{multline}
We denote the compact singular value decomposition (SVD) of $\vect{A}_g[s]$ as $\vect{A}_g[s] =\vect{U}_A[s]\vect{\Sigma}_{A}[s]\vect{V}_A^{\Htran}[s]$, where $\vect{\Sigma}_{A}[s]$ is the square matrix with the number of rows and columns being equal to the rank of $\vect{A}_g[s]$. Using this SVD representation of $\vect{A}_g[s]$, we express  $\hat{\vect{x}}_g[s]$ in \eqref{eq_pinv} as
\begin{align}\label{eq_pinv2}
  \hat{\vect{x}}_g[s] = \vect{V}_A[s]\vect{\Sigma}_A^{-1}[s]\vect{U}_A^{\Htran}[s] \left(\frac{1}{\sqrt{P}}\vect{y}[s]-\sqrt{\beta }e^{\imagunit\phi }\overline{\vect{H}}[s]\vect{a}(\varphi)  \right).
\end{align}
The objective function in \eqref{eq_MLE_xg} for the $s$-th term can be written as $\hat{\vect{x}}_g^{\Htran}[s] \vect{A}_g^{\Htran}[s] \vect{A}_g[s]  \hat{\vect{x}}_g[s]  - 2 \Re \left( \hat{\vect{x}}_g^{\Htran}[s] \vect{A}_g^{\Htran}[s] \vect{A}_g[s]  \hat{\vect{x}}_g[s]  \right)  = -   \hat{\vect{x}}_g^{\Htran}[s] \vect{A}_g^{\Htran}[s] \vect{A}_g[s]  \hat{\vect{x}}_g[s] $. 
Therefore, the MLE problem in \eqref{eq_MLE_4vars} can be rewritten as in  \eqref{eq_MLE_3vars} given at the top of the next page, 
where $\overline{\vect{A}}[s]=\vect{I}_M-\vect{U}_d\vect{U}_d^{\Htran}-\vect{U}_A[s]\vect{U}_A^{\Htran}[s]$. 
\begin{figure*}
\begin{align}\label{eq_MLE_3vars}
& \argmin{\beta ,\phi ,\varphi}  \sum_{s=1}^S \Bigg(     - \vect{y}^{\Htran}[s]\vect{U}_A[s]\vect{U}_A^{\Htran}[s]\vect{y}[s]  - P \beta   \vect{a}^{\Htran}(\varphi)  \overline{\vect{H}}^{\Htran}[s]\vect{U}_A[s]\vect{U}_A^{\Htran}[s] \overline{\vect{H}}[s] \vect{a}(\varphi) + 2  \sqrt{P}\sqrt{\beta } \Re \left(   e^{\imagunit \phi } \vect{y}^{\Htran}[s]\vect{U}_A[s]\vect{U}_A^{\Htran}[s] \overline{\vect{H}}[s] \vect{a}(\varphi) \right) 
\nonumber \\
& \hspace{6em}+ P {\beta}  \vect{a}^{\Htran}(\varphi) \overline{\vect{H}}^{\Htran}[s] (\vect{I}_M - \vect{U}_d\vect{U}_d^{\Htran}) \overline{\vect{H}}[s] \vect{a}(\varphi)     -  2\sqrt{P}\sqrt{\beta }  \Re{\left(   e^{\imagunit \phi  }  \vect{y}^{\Htran}[s] (\vect{I}_M-\vect{U}_d \vect{U}_d^{\Htran} )     \overline{\vect{H}}[s] \vect{a}(\varphi) \right)}\Bigg), 
 \nonumber \\
& =\argmin{\beta ,\phi ,\varphi} \quad \sum_{s=1}^S \Bigg(P {\beta}  \vect{a}^{\Htran}(\varphi) \overline{\vect{H}}^{\Htran}[s]  \overline{\vect{A}}[s] \overline{\vect{H}}[s] \vect{a}(\varphi) -  2 \sqrt{P} \sqrt{\beta }  \Re{\left(   e^{\imagunit \phi  }  \vect{y}^{\Htran}[s] \overline{\vect{A}}[s]     \overline{\vect{H}}[s] \vect{a}(\varphi) \right)}\Bigg).
\end{align}
\hrulefill
\end{figure*}

\begin{figure}
\centering
\subfloat[The row-to-column ratio of $\vect{U}_d$.]
{ \includegraphics[width=0.23\textwidth]{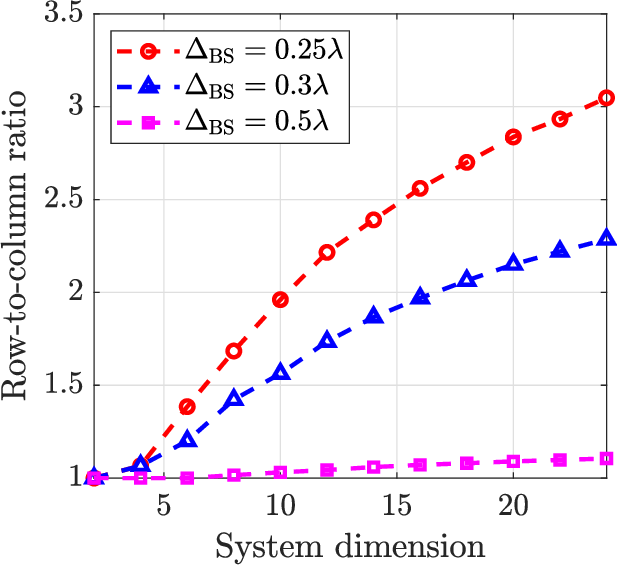}}\hfill
\centering
\subfloat[The number of effective eigenvalues of $\overline{\vect{A}}$.]
{ \includegraphics[width=0.23\textwidth]{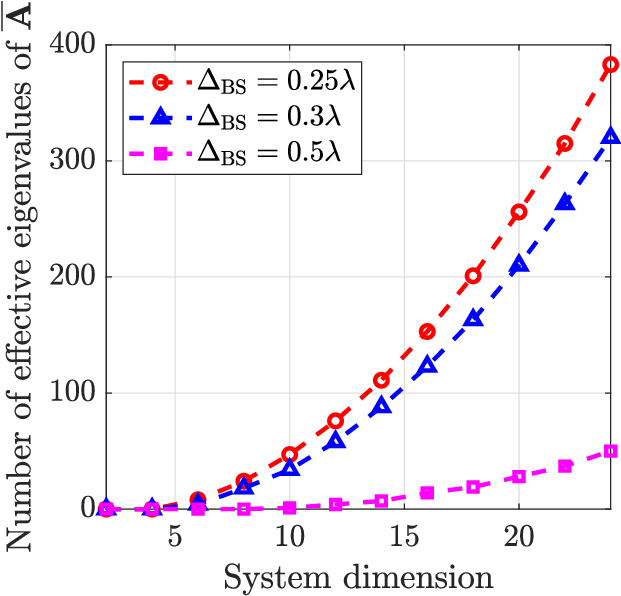}}
\caption{Evaluation of matrices $\vect{U}_d$ and $\overline{\vect{A}}[s]$ with $P = 15$ dBm and $\kappa  = 16$\,dB.}
\label{F_Ud_Ratio}
\end{figure}
From \eqref{eq_Ag}, we know that  $\vect{U}_A[s]$ resides in the subspace spanned by $\vect{I}_M-\vect{U}_d\vect{U}_d^{\Htran}$. It holds that $\overline{\vect{A}}[s]$ is a positive semidefinite matrix when $\vect{U}_d$ is a tall matrix obtained when the number of BS antennas  is large in a UPA geometry \cite{demir2022channel}. We evaluate the ratio of the numbers of rows-to-columns of the matrix $\vect{U}_d$ in Fig.~\ref{F_Ud_Ratio}(a). 
The size of the matrix $\vect{U}_d$ depends on the number of BS antennas since it denotes the subspace where the channel vector between the BS and UE exists. We define a system dimension to represent the number of antennas in each row or column of a UPA with a square geometry. From the figure, we see that increasing the system dimension as well as reducing the antenna spacing results in a taller matrix $\vect{U}_d$. In addition, we plot the number of effective eigenvalues of $\overline{\vect{A}}[s]$, i.e., the sum of non-zero positive eigenvalues of $\overline{\vect{A}}[s]$, in Fig.~\ref{F_Ud_Ratio}(b). We can see that the number of effective eigenvalues increases with system dimension. Therefore, we conclude that the MLE needs a sufficiently large number of BS antennas (e.g., $ M_{\rm H} = M_{\rm V} =10 $) to acquire a positive semidefinite matrix $\overline{\vect{A}}[s]$.
Without loss of generality, we consider an inter-antenna spacing of $\Delta_{\rm BS} = 0.25 \lambda$ at the BS and assume that the number of BS antennas is sufficiently large such that the matrices $\overline{\vect{A}}[s]$ are positive semidefinite. This latter condition will be utilized in the subsequent MLE steps.

It is important to note that the matrices $\vect{U}_d$ and $\vect{U}_g$ span all possible channel realizations. The rank deficiency of these matrices is solely attributed to the UPA array geometry, which results from oversampling with a UPA \cite{demir2022channel}.

We proceed to estimate $\phi $ minimizing the objective function in \eqref{eq_MLE_3vars}, where only the second term in the objective function depends on $\phi $. As this term has a negative sign, the minimization problem transforms into a maximization problem. Utilizing the Euler representation of $\sum_{s=1}^S\vect{y}^{\Htran}[s] \overline{\vect{A}}[s] \overline{\vect{H}}[s] \vect{a}(\varphi)$, we obtain the solution
\begin{equation}
    \hat{\phi}  = -\arg\left( \sum_{s=1}^S \vect{y}^{\Htran}[s] \overline{\vect{A}}[s] \overline{\vect{H}}[s] \vect{a}(\varphi) \right).
\end{equation}
which makes $\sum_{s=1}^S\Re{\left(   e^{\imagunit \phi  }  \vect{y}^{\Htran}[s] \overline{\vect{A}}[s]     \overline{\vect{H}}[s] \vect{a}(\varphi) \right)} = \left| \sum_{s=1}^S  \vect{y}^{\Htran}[s] \overline{\vect{A}}[s] \overline{\vect{H}}[s] \vect{a}(\varphi)\right|$. 

Therefore, we can rewrite \eqref{eq_MLE_3vars} as 
\begin{multline}\label{eq_MLE_2vars}
    \argmin{\beta ,\varphi} \quad      
 \sum_{s=1}^S \Bigg(   P {\beta}  \vect{a}^{\Htran}(\varphi) \overline{\vect{H}}^{\Htran}[s]  \overline{\vect{A}}[s] \overline{\vect{H}}[s] \vect{a}(\varphi) \Bigg) \\
    -  2 \sqrt{P} \sqrt{\beta }  \left\vert      \sum_{s=1}^S  \vect{y}^{\Htran}[s] \overline{\vect{A}}[s]     \overline{\vect{H}}[s] \vect{a}(\varphi) \right\vert ,
\end{multline}
which is a quadratic minimization function with respect to $\sqrt{\beta }$ since $\overline{\vect{A}}[s]$ is a positive semidefinite matrix. The latter holds true when having a sufficiently large BS array, as illustrated in Fig.~\ref{F_Ud_Ratio}(b).  Hence, we obtain an estimate of $\beta $ as
\begin{equation}\label{eq_beta_hat}
    \hat{\beta}  =  \frac{\left\vert  \sum_{s=1}^S     \vect{y}^{\Htran}[s] \overline{\vect{A}}[s]     \overline{\vect{H}}[s] \vect{a}(\varphi) \right\vert^2} {  P \left ( \sum_{s=1}^S \vect{a}^{\Htran}(\varphi) \overline{\vect{H}}^{\Htran}[s]  \overline{\vect{A}}[s] \overline{\vect{H}}[s] \vect{a}(\varphi)\right)^2 }.
\end{equation}
By substituting \eqref{eq_beta_hat} into  \eqref{eq_MLE_2vars}, we derive the MLE with respect to the AoA $\varphi$ as
\begin{equation}
     \argmin{\varphi} -    \frac{ \left\vert   \sum_{s=1}^S \vect{y}^{\Htran}[s] \overline{\vect{A}}[s]     \overline{\vect{H}}[s] \vect{a}(\varphi) \right\vert^2} {\sum_{s=1}^S \vect{a}^{\Htran}(\varphi) \overline{\vect{H}}^{\Htran}[s]  \overline{\vect{A}}[s] \overline{\vect{H}}[s] \vect{a}(\varphi) } . 
\end{equation}
Therefore, the AoA estimate is given as
\begin{align}\label{eq_phi_hat}
    \hat{\varphi} = \argmax{\varphi} \quad   \frac{ \left\vert  \sum_{s=1}^S  \vect{y}^{\Htran}[s] \overline{\vect{A}}[s]     \overline{\vect{H}}[s] \vect{a}(\varphi) \right\vert^2} {\sum_{s=1}^S\vect{a}^{\Htran}(\varphi) \overline{\vect{H}}^{\Htran}[s]  \overline{\vect{A}}[s] \overline{\vect{H}}[s] \vect{a}(\varphi)},
\end{align}
which can be solved numerically via a grid search. 
The summation over multiple subcarriers in \eqref{eq_phi_hat} demonstrates how we can utilize subcarrier diversity to improve angle estimate accuracy leading to a more accurate channel estimate.
\begin{remark}
    The AoA estimate derived in \eqref{eq_phi_hat} can be interpreted as a wideband generalization of the parametric MLE presented in \cite[Eq. (22)]{2022_Björnson_Asilomar} for narrowband RIS systems with no NLOS channel between the UE and the BS. In scenarios where NLOS channels characterized by the reduced subspace matrix $\vect{U}_g$ and $\vect{U}_d$, are absent, $\vect{\overline{A}}[s]$ reduces to an identity matrix. If $S=1$ and there are no NLOS channels, the estimate of AoA in \eqref{eq_phi_hat} will be identical to the one presented in \cite{{2022_Björnson_Asilomar}}. 
\end{remark}

\section{Simulation Results}
\label{S_Numel}

In this section, we evaluate the proposed parametric MLE by comparing it with the narrowband MLE (NB-MLE) \cite{2022_Björnson_Asilomar}, where  the parameters are estimated individually at each subcarrier since it was developed for a narrowband system that does not consider multiple subcarriers. Furthermore, we compare the proposed MLE with a baseline MLE approach that does not incorporate the NLOS estimator to demonstrate the significance of our NLOS estimator. The baseline estimator is referred to as the NLOS-Unaware MLE, obtained by setting $\bar{\vect{A}}[s]$ as an identity matrix in our proposed MLE.

As performance metric, we consider the normalized mean square error (NMSE) defined as 
\begin{equation}
    {\rm NMSE }= \frac{ \mathbb{E} \{\sum_s \|\hat{\vect{x}}_s - \vect{x}_s \|^2_2 \} }{ \mathbb{E}\{\|\sum_s  \vect{x}_s\|^2_2\}},
\end{equation}
where $\hat{\vect{x}}_s$ is the estimate of $\vect{x}_s$. We consider the same number of BS antennas and RIS elements $M_{\rm H}=N_{\rm H}=8$ and $M_{\rm V}=N_{\rm V}=16$. We set the wavelength to $\lambda = 0.1$\,m (i.e., $3$\,GHz).
There are $40$ NLOS multipath components originating from random azimuth angles within the range of  $[-4\pi/9,4\pi/9 ]$ and elevation angles within $[-2\pi/9,2\pi/9 ]$ of the nominal angles. The nominal angles are set as follows.
The BS is oriented towards the RIS, i.e., $\bar{\varphi}_{\rm BS}=0$. The RIS is positioned facing the BS at an azimuth angle of $\bar{\varphi}_{\rm RIS }= \pi/4$. The UE is located at an azimuth angle of $\varphi = \pi/6$ relative to the BS and $\varphi_{\rm UE }= \pi/3$ relative to the RIS. 
The BS, RIS, and UE are all considered to be at the same elevation angle of $0$.
This scenario is illustrated in Fig.~\ref{F_illus}.  The complex gain of each NLOS path is complex Gaussian distributed with the same variance corresponding to a uniform power delay profile. 
Furthermore, we set the channel gains between BS and RIS, BS and UE, and RIS and UE as $\bar{\beta}  = -80 +\kappa_{\rm BS-RIS}$ dB, $\dot{\beta} = -70 $ dB, and ${\beta} = -124+ \kappa_{\rm RIS-UE}$ dB, respectively. The notation $\kappa$ can be viewed as a K-factor in the Rician fading model specifying the LOS-to-NLOS power ratio. 
Unless otherwise mentioned, we set $\kappa_{\rm BS-RIS} = \kappa_{\rm RIS-UE} = 16$\,dB. The noise variance is $\sigma^2 = -124$\,dB. The simulations are performed across $5000$ channel realizations. 

\begin{figure}
\centering
\subfloat[Channel estimation.]
{ \includegraphics[width=0.48\textwidth]{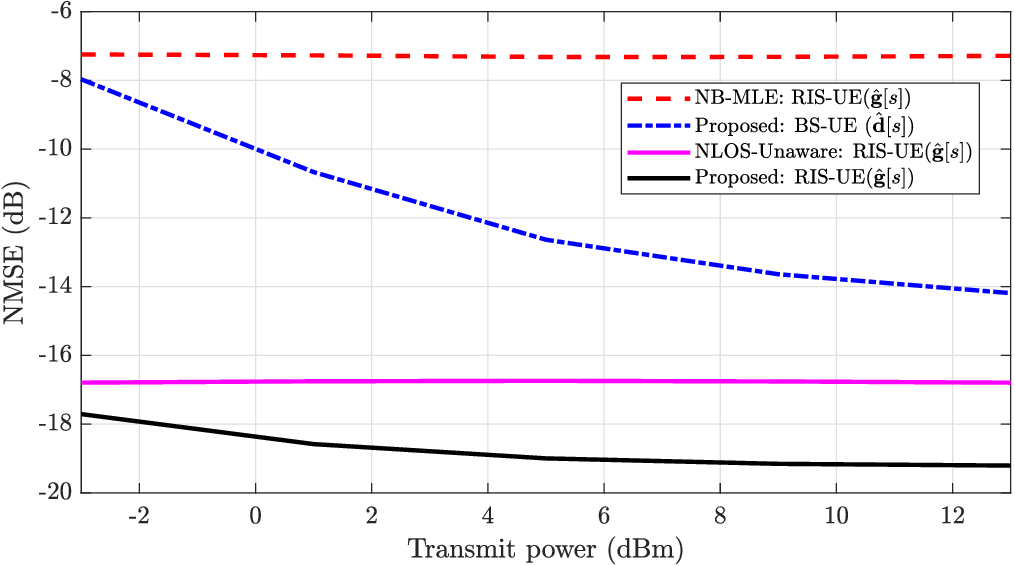}}\hfill
\centering
\subfloat[Angle estimation.]
{ \includegraphics[width=0.48\textwidth]{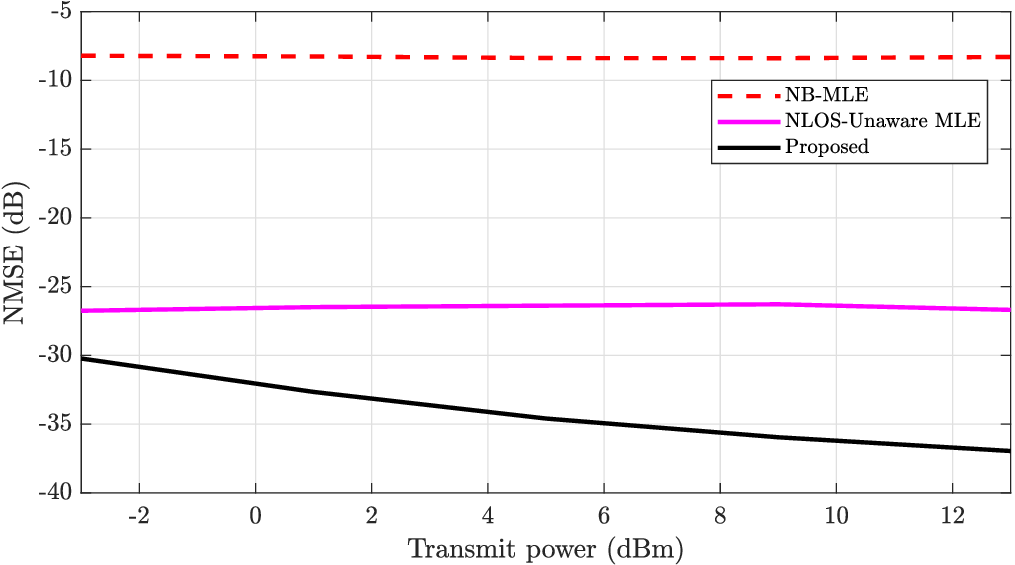}}
\caption{The NMSE with respect to the transmit power $P$.}
\label{F_NMSE_SNR}
\end{figure}

In Fig.~\ref{F_NMSE_SNR}, we evaluate the NMSE performance of the proposed MLE both for the BS-UE and RIS-UE channels. We use $S=16$ equally spaced subcarriers for pilot transmission. The performance of the proposed estimator significantly outperforms the NB-MLE when directly applied to our system model. Moreover, the channel estimation performance of the proposed MLE is around $2$\,dB better than the NLOS-Unaware MLE, as illustrated in Fig.~\ref{F_NMSE_SNR}(a). This highlights the substantial improvement achieved by taking the NLOS paths into consideration when deriving the MLE. We obtain similar findings  when evaluating the angle estimation performance of the considered estimators, shown in Fig.~\ref{F_NMSE_SNR}(b). 

\begin{figure}
\centering
\subfloat[$\kappa_{\rm BS-RIS} = \kappa_{\rm RIS-UE}. $]
{ \includegraphics[width=0.48\textwidth]{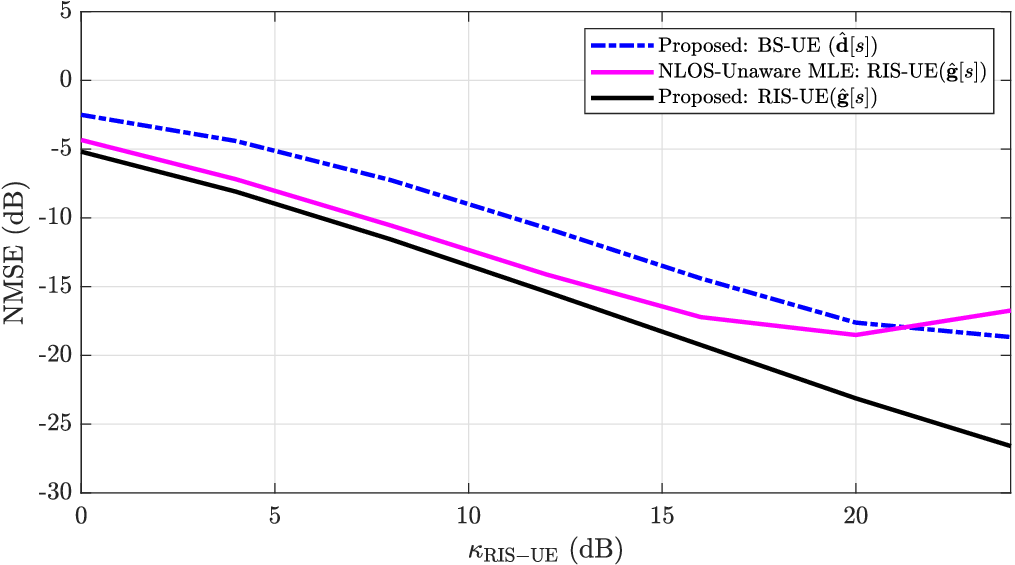}}\hfill
\centering
\subfloat[$\kappa_{\rm BS-RIS}=0$\,dB.]
{ \includegraphics[width=0.48\textwidth]{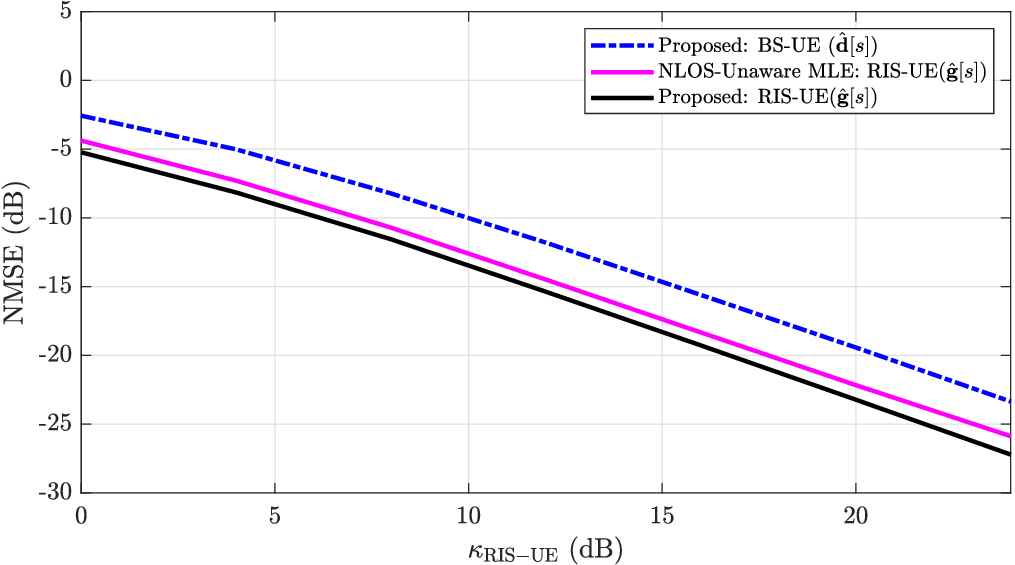}}
\caption{The NMSE with respect to the Rician K-factor  between the RIS and UE.}
\label{F_NMSE_kappa}
\end{figure}
In Fig.~\ref{F_NMSE_kappa}, we evaluate the NMSE performance of the proposed MLE with respect to the Rician K-factor. We set $P=15$\,dBm. As illustrated in Fig.~\ref{F_NMSE_kappa}(a), the NMSE of the proposed MLE decreases with the K-factor. Under dominant LOS conditions (e.g., $\kappa_{\rm BS-RIS}=\kappa_{\rm RIS-UE}=24$\,dB), the NMSE of the NLOS-Unaware MLE deteriorates, whereas the NMSE of the proposed MLE remains satisfactory. The performance degradation of the NLOS-Unaware MLE is due to the more pronounced low rank of the BS-RIS channel. To confirm this, we plot Fig.~\ref{F_NMSE_kappa}(b), the issue is resolved when $\kappa_{\rm BS-RIS}$ is fixed at $0$\,dB, indicating a rich scattering environment between the BS and RIS. Based on this observation, we conclude that our NLOS estimator not only enhances the performance of the proposed MLE but also makes it robust under rank-deficient BS-RIS scenarios.

\section{Conclusions}

We have proposed an efficient parametric MLE for channel estimation in RIS-assisted wideband systems. The MLE estimates the channels between RIS-UE and BS-UE based on their respective reduced-subspace channel representations. The MLE can be considered a generalized version of the existing MLE for narrowband RIS systems with no NLOS paths. The proposed estimator performed significantly better  than the NB-MLE and NLOS-unaware MLE when implementing in RIS-assisted wideband systems, as evidenced by achieving significantly lower NMSE.

\bibliographystyle{IEEEtran}
\bibliography{IEEEabrv,refs}

\end{document}